
%
%
%
%
%
%
\magnification=\magstep1
\def\bx{}
\def\sl{\rm }
\def\lin{{\bigg|}}
\def\RR{\hbox{\bf R}}
\def\CC{\hbox{\bf C}}

\def\tr{\hbox{\rm tr}}

\def\ch{\hbox{\rm ch}}
\def\ZZ{\hbox{\bf Z}}
\def\treq{\buildrel \rm tr \over =}
\def\defeq{\buildrel \rm def \over =}
\def\ts{\textstyle}

\def\ds{\displaystyle}
\def\exampleA{}
\def\exampleB{}
\def\proof{}
\font\title=cmr17
\font\ninerm=cmr9

\vfill\eject

\centerline{\title The algebra of Chern-Simons classes }

\centerline{\title  and the
 Poisson bracket on it.}\footnote{ }{\ninerm
 This work was partially
 supported by NSF grant DMS 92-13357
}

\vskip 20pt

\vskip 15pt
\centerline {\bf Israel M. Gelfand,}
\centerline {Mathematics Department, Rutgers University,}
\centerline { New Brunswick, N.J., 08903}
\vskip 5pt
\centerline {\bf Mikhail M. Smirnov,}
\centerline {
Mathematics Department, Princeton University,}
\centerline{ Princeton, N.J., 08544 smirnov@math.princeton.edu}
\vskip 20 pt
\centerline{{\bf Abstract}}
\medskip\noindent
{\ninerm
Developing  ideas of the formal geometry [G1], [GKF], and ideas based on
combinatorial formulas for characteristic classes
we introduce the algebraic structure  modeling
$N$ connections  on the vector bundle over an oriented manifold.
First we  construct  a graded free associative algebra $A$ with a differential
$d$.
Then we go to the space  $V$ of cyclic words of $A$. Certain elements of $V$
correspond
to the secondary characteristic classes associated to $k$ connections.
That construction allows us to  give easily the explicit formulas for some
known secondary classes and to construct the new ones. Space $V$ has  new
operations:
it is a graded Lie algebra with respect to the Poisson bracket. We write how
$i$-th
differential and $i$-th homotopy operator in the algebra are connected with
this bracket. There is an  analogy between our algebra and the Kontsevich
version of the noncommutative symplectic geometry. We consider then  an
algebraic model of the action of the gauge group. We describe how elements of
our algebra corresponding to the  secondary characteristic classes
change under this action.

}

\vskip 20pt\noindent
{\bf 0. Introduction.}
\vskip 10pt
\noindent
The problem  of
 localization of topological invariants by the methods of
formal geometry originated
in the combinatorial formulas for characteristic classes
(Gabrielov-Gelfand-Losik [GGL].
Closely related to  combinatorial formulas
for characteristic classes is the problem of explicit description of secondary
characteristic classes and difference cocycles  (Chern-Simons [ChS], Cheeger,
 Bott-Shulman-Stasheff [BSS], Youssin [You] and others).

The idea of   using the local data
(fields of geometric objects) and a
 construction of a universal field and a variational bicomplex
goes back to [G1] (see also the work of Gelfand, Kazhdan and Fuks [GKF]).

 We introduce here the algebraic structures which model a
certain differential geometric situation. Let $G$ be a Lie group with finitely
many components and $g$ be its Lie algebra (we shall take
$G=GL(n, \CC)$, or $GL(n,\RR)$ so $g$ consists of all $n\times n$
 matrices.
Let $X$ be a $C^\infty$ oriented manifold and let $E_G$ be a principal
 $G$ bundle over $X$ provided with $N$ connections $\nabla_1$, $\ldots$,
$\nabla_N$, ($N\neq n$). Let us associate to $E_G$ a
$n$-dimensional
  vector
bundle $\pi:E\to X$ and connections
$\omega_0(x)$, $\omega_1(x)$, $\ldots$, $\omega_N(x)$ on this bundle
given by their matrices of 1-forms
$\omega_0(x)$, $\omega_1(x)$, $\ldots$, $\omega_N(x)$
 $\in
\Omega^1(X)\otimes g$.
Let $\Delta_I$ be a $k$-dimensional simplex
with vertices $i_0, \ldots, i_k$ $\subset \{0,1,\ldots, N\}$,
$\Delta_I\cong \{ t_0+t_1+\ldots+t_k=1,$ $ t_0,\ldots,
t_k>0\}$.
Consider a connection $\omega(t)$, $t\in \Delta_I$ such that
$\omega(t)=t_0\omega_{i_0}+\ldots+t_k\omega_{i_k}$. Then we define a
secondary characteristic class
     $$\ch_m^k(\omega_{i_0}\ldots \omega_{i_k})=\tr\int_\Delta
       [d(t_0\omega_{i_0}+\ldots+t_k\omega_{i_k})+(t_0\omega_{i_0}+\ldots
+t_k\omega_{i_k})^2]^m,$$ where $d$ is the total differential (with respect to
$x$ and $t$ on $\Delta_I
\times X$ [GGL]) and in the expression under the integral we take
 only summands with $k$ $dt$'s and
forget about the other summands.

Then the following relation holds:
$$\ch_m^k(\omega_{i_0}\ldots \omega_{i_k})=
\sum_{p=1}^k(-1)^p\ch_m^{k-1}(\omega_{i_0}\ldots \hat{\omega}_{i_p}
\ldots\omega_{i_k}).$$
We are interested in the explicit expressions for $\ch_m^k$ and
the relations and algebraic operations on them.

In the works of Gabrielov, Gelfand  and Losik [GGL] and  Gabrielov, Gelfand
and Fuks
[GGF] two double complexes  were constructed. One was called the
variational
bicomplex and the other was called the difference bicomplex.

On the other hand Shulman [S], Bott [B] and Bott, Shulman
and Stasheff [BSS] in slightly different context
constructed a bicomplex very similar to the
difference bicomplex.  Secondary classes
appear in a natural way in these complexes.

It is known that the Chern-Simons class constructed from two connections $A_0$
 and $A_1$ has the form $\ch^1_2(A_0,
 A_1)$ $={1\over 2}\tr\big( (A_1 dA_1 + {2\over 3}A_1^3)-
 (A_0 dA_0 + {2\over 3}A_0^3) +d(A_1 A_0)\big),$
where $A_i$ are matrices of 1-forms which define connections.
For higher secondary classes the calculations give longer  polynomials.
For example
\vskip 5pt\noindent
\leftline{$\ch^1_3(A_0,
 A_1)$ $={1\over 2}\tr[ ({1\over 5}a_1^5+{1\over 3}
a_1 b_1 b_1+{1\over 2}a_1^3 b_1)- $ $ ({1\over 5}a_0^5+{1\over 3}
a_0 b_0 b_0+{1\over 2}a_0^3 b_0)+ $}
\vskip 5pt\noindent
\rightline{$+{1\over 6}d\big((a_0a_1-a_1a_0)(b_0+b_1)-(a_0^2+a_1^2+{1\over
2}a_1a_0)a_1a_0\big) ],$}
\vskip 5pt\noindent
\leftline{where $a_i=A_i$ and $b_i=dA_i$.}
So, even in the simplest case for normal secondary characteristic classes
 formulas are lengthy
and we have to handle
combinatorics of long noncommutative expressions.
 In this chapter we introduce algebraic structures which are useful in
calculations, but
 which are also of independent interest.
When we translate the situation to the algebraic language we can see
certain new things which are hard to see on the differential-geometric side.
Namely there is a remarkable parallel between the algebraic structure
which is described below and the Kontsevich version of the noncommutative
symplectic geometry [Kon]. We hope to return to this parallel later.
 Vaguely we can view the algebra $A$ described in the next section as a
'cotangent bundle' of its subalgebra  generated by $a$'s.

Now let us draw a parallel between the algebraic and the geometric pictures.
\vskip 10pt\noindent
$$\matrix{\hbox{GEOMETRY}&\hbox{ALGEBRA}\cr
&\cr
&\cr
\hbox{connections} \ \ \omega_i&\hbox{
generators} \ \ a_i\cr
&\hbox{ of the free algebra} \ A\cr
&\cr
d\omega_i&\hbox{
generators} \ \ b_i\cr
&\hbox{ of the free algebra} \ A\cr
&\cr
\tr & \hbox{Space of cyclic words \ } V\cr
&\cr
d&d\cr
&\cr
\hbox{ Curvatures \ } R_i=d\omega_i+\omega_i^2& (b_i+a_i^2)\in A\cr
&\cr
\hbox{Chern forms}&\cr
\ch_m(\omega_i)=\tr R_i^m&(b_i+a_i^2)^m\in V\cr
&\cr
\hbox{Gauge transformation} \
\sigma, \sigma^{-1}&\hbox{Generators} \ x, y \cr
&\hbox{in the gauge algebra}\ \ G \cr
&\cr
d\sigma \sigma^{-1}& c\cr
&\cr
\omega \to \sigma^{-1}d  \sigma +   \sigma^{-1} \omega  \sigma&
a  \to y(a+c)x\cr
&\cr
-&\partial_i\cr
&\cr
 - &\hbox{Poisson bracket} \{ \ , \ \}\cr
&\cr
\hbox{etc.}&\hbox{etc.}}$$
Apparently there are connections between the algebra introduced here and
BRST [LZ] which we do not discuss here.
We want to thank  V.Retakh, A.Gabrielov,  M.Kontsevich and R.MacPherson
for various discussions.

\vskip 10pt\noindent
{\bf 1. The associative algebra and its cyclic space.}
\vskip 10pt\noindent
Consider an associative noncommutative algebra $A$ over $\CC$ freely
generated by $N$ generators $$a_i, \ i=1,...,N, \ \hbox{\rm of degree 1}$$ and
$N$
generators $$b_i, \ i=1,...,N, \ \hbox{\rm of degree 2.}$$
Consider an operation $d$ defined on generators by $$ {da_i=b_i, \
i=1,...,N;}$$ $${ db_i=0, \ i=1,...,N;}$$
which is extended to monomials in $A$ by the rule
$d(PQ)=(dP)Q + (-1)^{ \vert P \vert} PdQ $ and then to
 $A$ by linearity. Here $P$ and $Q$
are  monomials in $a_i, \  b_i$ and $  \vert P  \vert $ is the degree of $P$
i.e.
the
sum of degrees of all letters in $P$.

\vskip 5pt\noindent
{\bf Example 1.} Let $A$ be generated by 2 letters $a$
and $b$. Then the elements of $A$ are noncommutative polynomials in
$a$ and $b$. For example,
  $L=5aaba + aaa + 3bb$ is an element of $A$. Then $dL=5((da) aba $ $+ (-1)^1 a
(da)ba
+  $ $(-1)^2 aa (db)a + (-1)^4 aab(da))$ $ +((da)aa+(-1)a(da)a +(-1)^2aa(da))
+3((db)  \cdot b+(-1)^2b (db))$ $=5(baba-abba+aabb)+(baa-abaaab).$

\vskip 5pt\noindent
{\bf Definition 1.} Consider the factor space
$V=A/\{PQ-(-1) ^{ \vert P  \vert  \vert Q  \vert} QP\}$, where
$\{PQ-(-1)^{  \vert P  \vert  \vert Q  \vert } QP\}$ is the subspace of $A$
spanned
by the commutators of all monomials in  $A$. The space $V$ is generated by
cyclic words in letters $a_i, \  b_i$ and we shall call $V$ a cyclic space. We
are going to use symbol  $\treq$ for equality in $V$ to distinguish
it from the equality in $A$ which we  denote by $=$ \ .

\vskip 5pt\noindent
{\bf Example 2.} Consider a cyclic space $V$ of the algebra $A$
generated by two letters $a$ and $b$. Let us write its basis:
\vskip 5pt\noindent
$a, \ ( a^2  \treq 0),  \ b, \ ba  \treq ab,$ $ \ a^3, \ (a^4  \treq 0),$
 $ a^2b  \treq ba^2
 \treq -aba,$ $b^2,$ $ a^5, $ $ a^3b  \treq a^2ba $ $ \treq aba^2  \treq ba^3,$
 $ b^2a$
$ \treq ab^2,$  $(a^6  \treq 0)$, $a^4b  \treq -a^3ba  \treq  \ldots   \treq
ba^4,$
$b^2a^2  \treq -abba  \treq a^2b^2,$ $ abab  \treq baba  \treq 0,$ etc.
\vskip 5pt\noindent
So we have the basic monomials and all other monomials are obtained by a cyclic
permutation with an appropriate sign from the basic monomials:
$a$, $b$, $ab$, $a^3,$ $a^2b$, $b^2$, $a,$ $ a^3b$, $ab^2$, $a^4b,$ $b^2a^2$,
$\ldots $

\vskip 5pt\noindent
{\bf Example 3.} In the cyclic space $V$ of the algebra $A$ generated
by $4$ letters $a_1,a_2$ and $b_1,b_2$,

  $a_1^2  \treq 0$ because $a_1 a_1  \treq (-1)^{1\cdot 3} a_1 a_1  \treq
-a_1 a_1$.

  $a_1^4  \treq 0$ because $a_1^4  \treq a_1 a_1^3  \treq (-1)^{1\cdot 3}
a_1^3a_1  \treq -a_1^4$. More  generally

$a_i^{2k}  \treq 0$.

$a_1b_2a_1a_1a_2b_1  \treq (-1)^{1  \cdot 7} b_2a_1a_1a_2b_1a_1\treq$
$ (-1)^{2  \cdot 6} a_1a_1a_2b_1a_1b_2\treq$

 $\treq (-1)^ {1\cdot 7} a_1a_2b_1a_1b_2a_1\treq $
etc.$\treq $ $(-1)^{2  \cdot 6}$ $b_1a_1b_2a_1a_1a_2$.

\vskip 5pt\noindent
{\bf Proposition 1.}
\vskip 0pt
\noindent
1. The operator $d$ is a differential in the associative
algebra $A$: \  $d  \circ d=0$.
\vskip 0pt
\noindent
2. The restriction of $d$ to the cyclic space $V$ is a well-defined
differential and $d  \circ d   \treq 0$ in $V$.
 If  $P  \treq 0$,  than $dP  \treq 0$.

\noindent\bx

\vskip 5pt\noindent
{\bf Example 4.} $a^2  \treq 0$ because $a  \cdot a=(-1)^{1  \cdot 1}
a  \cdot a$,
$da^2=ba-ab  \treq ba-ba=0$.

\vskip 10pt\noindent
{\bf 2. Symmetric functions.}
\vskip 10pt
 To write explicit formulas for
secondary  characteristic classes we  need the special class of
symmetric functions in noncommuting variables.

\vskip 5pt\noindent
{\bf Definition 1.} A symmetric function $ \Sigma [P^k,Q^l]$ in $A$
is a sum of all possible words with $k$ letters $P$ and $l$ letters $Q$.
\vskip 5pt\noindent
{\bf Example 1.}  $ \Sigma [P^1,Q^2]=PQQ+QPQ+QQP$.
We can replace $P$ and $Q$ by any elements of $A$, for example
$$  \Sigma [(a^2)^1,(b)^2]= a^2bb+ba^2b+bba^2$$
$$   \Sigma [(a^2+b)^1,(ab)^2]=(a^2+b)(ab)(ab)+(ab)(a^2+b)(ab)+
(ab)(ab)(a^2+b)$$
\vskip 5pt\noindent
{\bf Definition 2.} A symmetric function $ \Sigma [P_1^{k_1},\ldots
,P_m^{k_m}]$ in $A$
is a sum of all possible words with $k_1$ letters $P_1$, $k_2$ letters $P_2$
, $\ldots$, $k_m$ letters $P_m$.
\vskip 5pt\noindent
{\bf Example 2.}  $ \Sigma [P^1,Q^1,R^1]=PQR+RPQ+QRP+QPR+PRQ+RQP,$
$$ \Sigma[P^2,Q^1,R^1]=PPQR+PPRQ+PQPR+PRPQ+PQRP+PRQP+$$
\rightline{$+QPPR+RPPQ+QPRP+RPQP
+QRPP+RQPP.$}
\vskip 5pt\noindent
We can see that it is not important in which order the arguments
$P$, $Q$, etc. stay in brackets, what matters is only the exponents
which show how many  times we should take the same letters.
\vskip 5pt\noindent
{\bf Example 3.} Let $\Sigma_{p,q}=\Sigma[(a^2)^p, b^q]$
 be the sum of all possible words with $p$ letters $a^2$ and $q$
letters $b$.

$  \Sigma_{3,1}=bbba^2+bba^2b+ba^2bb+a^2bbb \treq 4bbba^2$

$  \Sigma_{3,2}  \treq 5bbba^2a^2+5bba^2ba^2$

$  \Sigma_{3,3}  \treq 6bbba^2a^2a^2 + 6bba^2ba^2a^2 + 6bb2a^2ba^2 +
2ba^2ba^2ba^2$



\vskip 10pt\noindent
{\bf 3. Other operations in the algebra, Poisson bracket.}
\vskip 10pt \noindent
The remarkable fact is that $V$ possess a Lie algebra structure with respect to
a Poisson bracket.
So we have  a
 Poisson bracket on the space containing characteristic and secondary
characteristic
classes.

First we define certain new operations in $A$. Let $A$
as before be
a free associative algebra
generated by $ a_i$ and $b_i,$ $ i=1,  \ldots , N. $
\vskip 5pt\noindent
{\bf Definition 1.} Partial derivative
$ { \partial  \over  \partial z}$ or $ \partial_z$, where
  $ z=a_i$ or $b_i$ is defined
on monomials by the following rule:
\vskip 3pt
\noindent 1. If the letter $z$
does not appear  in
the monomial $P$, then $  \partial_z P=0$.
\vskip 3pt
\noindent
2. If $P=A_1z A_2z A_3z  \ldots zA_k$, where $A_i$ are words
(may be empty) without the letter $z$
then $$  \partial_z P= (-1)^{ \varepsilon_1}/\!\!\! z A_2 zA_3  \ldots  A_k
A_1+
(-1)^{ \varepsilon_2}/\!\!\! z A_3z
 \ldots  A_k A_1z A_2+ \ldots$$ $$+  (-1)^{ \varepsilon_k}/\!\!\! z
A_k A_1z A_2  \ldots A_{k-1}. $$

 $  \varepsilon_1= \vert A_1  \vert  \vert zA_2  \ldots  A_k  \vert,$

$ \varepsilon_2= \vert A_1z A_2  \vert  \vert zA_3  \ldots  A_k  \vert $,
 etc.
\vskip 3pt \noindent
We take the initial monomial $P$ and look at all appearences of  $z$ in it.
Then for every letter $z$ we  cyclically permute the word
so that this letter $z$ becomes the first letter of the permuted word.
And then we delete this letter $z$. The sum of resulting words will
be a partial derivative in $z$.

\vskip 5pt\noindent
{\bf Example 1.} Suppose $A$ is generated by two letters $a$ and $b$.
Then $  \partial_a (aaabab)=  /\!\!\!aaabab+(-1)^{1  \cdot 7}/\!\!\!
 aabab a $ $
(-1)^{2  \cdot 6}/\!\!\! ababaa+(-1)^{5  \cdot 3}$ $ /\!\!\!a b
 aaab=aabab-ababa+babaa-baaab$

$  \partial_b(aaabab)=(-1)^{3  \cdot 5}/\!\!\! babaaa+(-1)^{2  \cdot 6}
/\!\!\!baaaba=
aaaba-abaaa.$
\vskip 5pt\noindent
{\bf Proposition 1.} The partial derivative is well-defined in the space
of cyclic words $V$.
That means that if $L=PQ$ is a representative in $A$ of a cyclic word
$L  \in V$ and $L'  \treq L$ another representative in $A$ of the same
 cyclic word
 so that $L'=(-1)^{ \vert P  \vert  \vert Q  \vert}QP$, then
$  \partial_z L  \treq  \partial_z L'$, \ { i.e.}
$$\partial_z (PQ)  \treq
(-1)^{ \vert P  \vert  \vert Q  \vert } \partial_z (QP) $$

\vskip 5pt\noindent
{\bf Definition 2.} Let $P$ and $Q$ be elements of $V$. Then their
Poisson bracket is $$ \{ P,Q  \}\treq  \sum\limits^N_{i=1}
(  \partial_{a_i} P  \cdot
 \partial_{b_i}Q+(-1)^{ \vert P  \vert \vert Q \vert
}  \partial_{a_i} Q
 \cdot  \partial_{b_i}P).$$
 Elements $P$ and $Q$ are in $V$.
 We take any of their preimages in $A$ also denoted by $P$ and $Q$
and  apply to them
  operators $\partial_{a_i}$
and $\partial_{b_i}$  defined in $A$. Then
 $ \partial_{a_i}P$ and $  \partial_{b_j}Q$
are elements of $A$. We multiply them in $A$
 and only then we take the corresponding cyclic element which will be
independent
on decyclization.
\vskip 5pt\noindent
{\bf Theorem 1.} Poisson bracket is well defined on $V$ and is linear in
its arguments.
It has properties
\vskip 5pt\noindent
1. $|\{P, Q\}|=|P|+|Q|-3,$ where $|P|$ is a degree of $P$.
\vskip 5pt\noindent
2. $ \{ P,Q  \}\treq (-1)^{ \vert P  \vert  \vert Q  \vert}  \{ Q,P  \}$.
\vskip 5pt\noindent
3. $ \{ P,  \{ Q,R  \}  \}+ (-1)^{ \vert P  \vert (  \vert Q  \vert +  \vert R
 \vert)}  \{ Q,  \{ R,P  \}
 \} $ $+(-1)^{ \vert R  \vert  \vert \{ Q,   P\}
 \vert}  \{ R,  \{ P,Q  \}  \} \treq 0$.
\vskip 5pt\noindent
 ($\ZZ_2$-graded Jacobi identity).
\proof
\vskip 5pt\noindent
{\bf Definition 3.} Define the operator   $d_i$ by the rule
$d_i a_i=b_i$, $ d_i b_i=0$, $d_i a_j=0$, $d_i b_j=0$,
and for a monomial $P=x_1\ldots x_p\in A$
$$d_iP=\sum_{k=1}^p (-1)^{|x_1\ldots x_{k-1}|}(x_1\ldots x_{k-1})(d_ix_k)
(x_{k+1}\ldots x_p),$$
Then $d_i$ is just a part of the differential $d$ which acts only
on the letters with index $i$: $d=\displaystyle{\sum_id_i}$.
\vskip 5pt\noindent
 {\bf Proposition 2.}
\vskip 0pt\noindent
1. The operator $d_i$ in the cyclic space $V$ can be written as the
bracket with
${1  \over 2} b_i^2$ and operator $d$  can be written as the bracket with
${1  \over 2} (b_1^2+  \ldots b_N^2)$;
$$
dP  \treq \textstyle{1  \over 2 }  \{ (b_1^2+  \ldots +b_n^2), P  \}. $$
\vskip 0pt\noindent
2.  Operators $d_i$, $d$ and the Poisson bracket are connected by equality
$$  -d_i  \{ P,Q  \}\treq  \{ d_iP, Q  \} + (-1)^{ \vert P  \vert}  \{ P, d_i
Q  \}, $$
$$  -d  \{ P,Q  \} \treq \{ dP, Q  \} + (-1)^{ \vert P  \vert}  \{ P, dQ  \}.
$$
\vskip 5pt\noindent
{\bf Proof.} Property 1 immediately follows from the definition
of the Poisson bracket and the fact that
$$dP  \treq  \sum\limits^N_{i=1} b_i (  \partial_{a_i} P).$$
Property 2 follows from the Jacobi identify for
$$R= \textstyle{1  \over 2} (b_1^2+ \ldots +b_n^2). \bx $$
The alternative proof can be done using the following lemma:

 \vskip 5pt\noindent
{\bf Lemma 1.} {\sl \  The following equalities are true  in $A$}
 $$\partial_{a_i}\circ d_i=d_i\circ \partial_{a_i},$$
$$\partial_{b_i}\circ d_i=d_i\circ \partial_{b_i}+\partial_{a_i}.$$
\proof
\vskip 5pt\noindent
{\bf Proposition 3.} We can express partial derivatives through
Poisson bracket as
$$  \partial_{b_i} P\treq  \{ a_i, \ P  \},  \ \ \ \ \ \
 \partial_{a_i}P\treq \{ P, \ b_i  \}.
$$
\bx
\vskip 5pt\noindent
{\bf Proposition 4.} Define operator $ d^k $ on $V$ by equality
$$d^kP  \treq  \sum\limits^N_{i=1} b^k_i (  \partial_{a_i} P). $$
Then
\vskip 3pt\noindent
1) $d^k  \circ d^k =0$, besides $d^1=d$, where $d$ is our standard differential
on $V$.
\vskip 5pt\noindent
2) The operator $d^k$ can be written as a Poisson bracket with
$ {1  \over {k+1}} (b_1^{k+1}+  \ldots +b_N^{k+1}):$
$$
d^kP  \treq \textstyle{1  \over {k+1}}  \{ (b_1^{k+1}+
 \ldots + b_N^{k+1}),P  \}$$
\vskip 5pt\noindent
{\bf Proof.}  Same as Proposition 3.
\bx
\vskip 10pt\noindent
{\bf 4. Chern characters. }
\vskip 10pt\noindent
{\bf Definition 1}
Consider now the algebra $A$ generated by two letters $a$ and $b$ and
cyclic image in $V$ of polynomials $ {1  \over {k!}} (a^2+b)^k $
Let us call corresponding elements of $V$ the $k-$th Chern character
 and denote
them by $\ch_k$. So $ \ch_k  \treq (a^2+b)^k $.

\vskip 5pt\noindent
We introduced  in $A$  symmetric functions
 of noncommuting variables $\displaystyle{\Sigma_{p, q}}=$ \
sum of all possible words
with $p$ letters $a^2$  and $q$ letters $b$.
For example, $\displaystyle{\Sigma_{1, 2}}=a^2bb+ba^2b+bba^2$.

 \vskip 5pt\noindent
{\bf Theorem  1.\ }
 {\sl There is a correspondence between ordinary and secondary characteristic
classes
and certain elements of $V$. The element $(b+a^2)^k$
 corresponds to a Chern character $\ch_k$ and
its differential is $0$:  $$d\big((b+a^2)^k\big)\treq  0.$$
Moreover
$$d\big(\textstyle{1\over
{(k-1)!}}a\big(\textstyle{1\over{k}}b^{k-1}+\textstyle{1\over{k+1}}{\Sigma_{1,
k-2}}
+\textstyle{1\over{k+2}}{\Sigma_{2, k-3}}+\ldots
+\textstyle{1\over{2k-1}}a^{2(k-1)}\big)\big)
\treq {1\over{k!}} (b+a^2)^k$$ and the expression in brackets on
 the left hand side as an element
of $V$  corresponds
 to a secondary class  $\ch^1_k$.}
\proof

\vskip 5pt\noindent
 {\bf Corollary.} $d  \Sigma_{p,q}=d  \Sigma[(a^2)^p,(b)^q]  \treq 0.$
\vskip 5pt\noindent
\vskip 5pt\noindent
In the following example we shall write where convenient $\underbrace
{b\ldots b}_{k \ \  times}$ for $b^k$ etc.
\vskip 5pt\noindent
{\bf Example1.}
To translate these formulas to the physics notation
we should take $a=A$ matrix of 1-forms defining the connection
take $b=dA$ matrix of two forms and put $Tr$ in front of the expression.
\vskip 5pt\noindent
\leftline{1. $k=2$}
 \vskip 5pt\noindent
$\ch_2^1\treq {1  \over 1!} a({1  \over 2}b+{1  \over 3} a^2)\treq {1  \over 2}
(ab+ {2  \over 3}a^3),$ \  \  in standard notation \ \ ${1  \over 2}Tr
(AdA+ {2  \over 3}A^3),$
\vskip 5pt\noindent
$ d\ch_2^1  \treq {1  \over 2}(a^2+b)^2\treq\ch_2^0$
\vskip 5pt\noindent
 \leftline {2. $k=3$}
 \vskip 5pt\noindent
$\ch_3^1\treq{1  \over 2!} a ({1  \over 3}b^2+{1  \over 4} (a^2b+ba^2)+
{1  \over 5}a^4)\treq{1  \over 3!}(ab^2+ {3  \over 2}a^3b+{3  \over 5} a^5)$
 \vskip 5pt\noindent
$d \ch_3^1  \treq {1  \over 3!}(a^2+b)^3\treq\ch_3^0$
\vskip 5pt\noindent
\leftline {3. $k=4$}
 \vskip 5pt\noindent
$\ch_4^1\treq{1  \over 3!} a( {1  \over 4}b^3+ {1  \over 5}(a^2bb+ba^2b+bba^2)+
{1  \over 6}(a^2a^2b+a^2ba^2+ba^2a^2)+{1  \over 7} a^6)\treq$
\vskip 2pt
{$ \treq{1  \over 4!}
(ab^3+{4  \over 5}(a^3bb+aba^2b+abba^2)+{4  \over 6}(a^5b+a^3ba^2+aba^4)
+{4  \over 7} a^7)\treq$}
\vskip 2pt
 \rightline{$\treq{1  \over 4!}(ab^3+{8  \over 5} a^3bb+ {4  \over 5}
aba^2b+2a^5b+{4  \over 7} a^7).$}

$d \ch_4^1\treq{1  \over 4!} (a^2+b)^4\treq\ch_4^0$
\vskip 4pt\noindent
 \leftline {4. $k=5$}
\vskip 5pt\noindent
$\ch_5^1\treq{1  \over 4!} a({1  \over 5}b^4+{1  \over 6}(a^2bbb+ba^2bb+bba^2b
+bbba^2)+$
\vskip 4pt

$+{1  \over 7}(a^2a^2bb+a^2ba^2b+ba^2a^2b+a^2bba^2+ba^2ba^2+bba^2a^2)
+$
\vskip 4pt\noindent
\rightline{$+{1  \over 8}(a^2a^2a^2b+a^2a^2ba^2
+a^2ba^2a^2+ba^2a^2a^2)+{1  \over 9} a^8)$}
\vskip 5pt\noindent
$\treq{1  \over 5!}(ab^4+{25  \over 6} a^3bbb+ {5  \over6} aba^2bb + {5  \over
6}
abba^2b +{{3  \cdot 5}  \over 7} a^5bb+{{2  \cdot5}  \over 7} a^3ba^2b +$
\vskip 5pt\noindent
\centerline{$+{5  \over 7} aba^4b
 + {{4  \cdot 5}  \over 8} a^5b + {5  \over 9} a^8).$}
\vskip 5pt\noindent
\vskip 5pt\noindent
There are also other  identities in this algebra.
\vskip 0pt\noindent
 Let
$ S_{ \alpha_1  \alpha_2 \ldots   \alpha_k }\treq$ $ ba^{  \alpha_1} ba^{
\alpha_2}b
\ldots ba^{  \alpha_k}$. For
example,
$ S_{0,3}= bbaaa$,   $ S_{1,2,0}=babaab$
\vskip 5pt\noindent
{\bf Proposition 1.}
\vskip 5pt\noindent
$ \textstyle{k  \over {2k-1}} d(a^{2k-1})\treq  \Sigma_{1,k-1}$
\vskip 5pt\noindent
$ \textstyle\textstyle{2  \over {2k+1}} d(ba^{2k-1})\treq  \Sigma_{2,k-2}$
\vskip 5pt\noindent
$\textstyle{k\over{2k-3}}d[(k-1)S_{0,2k-5}+1S_{1,2k-6}+(k-2)S_{2,2k-7}+2S_{3,2k-8}+
\ldots]\treq {\Sigma_{k-3, 3}}.$
\proof
\vskip 5pt\noindent
{\bf Example 2.} Let  $S_{p,q}=ba^pba^q$

$  \Sigma_{3,2} \treq 5bbba^2a^2 + 5bba^2ba^2\treq {5  \over 7} d( 3  \cdot
bba^5 + 1  \cdot baba^4 +
2ba^2ba^3)$

$  \Sigma_{3,3}  \treq {6  \over 9} d(4  \cdot bba^7 + 1  \cdot baba^6 +
3ba^2ba^5 + 2ba^3ba^4)$

$  \Sigma_{3,4} \treq {7  \over 11} d(5   bb  a^9 + 1 baba^8
+4ba^2ba^7 + 2ba^3ba^6 +3ba^4ba^5)$

$  \Sigma_{3,5}\treq {8  \over 13} d(6  \cdot S_{0,11} + 1  \cdot S_{1,10} +
5  \cdot S_{2,9} + 2\cdot S_{3,8} + 4  \cdot S_{4,7} + 3  \cdot S_{5,6})$

\vskip 10pt\noindent
{\bf 5. Algebra with dependence on $t$ and  secondary characteristic classes.}
\vskip 10pt
 Let $A$ be a free associative algebra generated by
elements $a(t)$, $\dot{a}(t)$ of degree 1 and elements $b(t)$, $\dot{b}(t)$
of degree 2, where $t\in [0,1]$. This algebra has a differential
$d$ such that
\vskip 5pt\noindent
$da(t)=b(t), \ \ \  d \dot{a}(t)=\dot{b}(t),$ \ \ \  $db(t)=0,
\ \ \ d\dot{b}=0.$
\vskip 5pt\noindent
This algebra has also a differential $\delta=d_t$ such that
\vskip 5pt\noindent
$d_t a(t)= dt \,\dot{a}(t) \  $  \ and  \ $d_t b(t)= dt\, \dot{b}(t)  $,
\vskip 5pt\noindent
and a derivative in $t$ for  expressions in $a, b$.
We denote this derivative by the dot.
We are not going to apply $d_t$ to expressions with $\dot{a}$ and
$\dot{b}$ (in other cases we should be introducing
letters with more dots but we don't want to deal with them).

\vskip 5pt\noindent
 From this algebra, as before, we construct  a factor space
 $V=A/\{ CB-(-1)^{|C||B|}BC \}$ where
$\{ CB-(-1)^{|C||B|}BC \}$ is the subspace of $A$ generated by commutators. We
call
$V$ a space of cyclic words and we denote by $\treq$ the equality in $V$.

Consider a path $a(t)$, $0\leq t\leq 1$ connecting $a_0$ and $a_1$and consider
$$\ch_k^1(a(t))\defeq\int_0^1 [(d_t+d)a(t)+a(t)^2]^k.$$
Now we can write an explicit formula for $\ch_k^1(a(t))$.

We shall use  certain  symmetric functions
 of noncommuting variables:

\noindent $\displaystyle{\Sigma_{p, q}}=$
 $\Sigma[{(a^2)^p,
(b)^q}]
=$sum of all possible words
with $p$ letters $a^2$  and $q$ letters $b$.
For example, $\displaystyle{\Sigma_{1, 2}}=a^2bb+ba^2b+bba^2$.
Also we shall use  functions
$\Sigma_{\dot{1},p,q}=\Sigma[{\dot{a},(a^2)^p,
(b)^{q}}]=$sum of all possible words with 1 letter $\dot{a}$
with $p$ letters $a^2$  and $q$ letters $b$.

\vskip 5pt\noindent
{\bf Theorem 1.} {\sl The secondary characteristic class defined by a path
$a(t)$
$$\ch^1_k(a(t))\treq
 {\ts{1\over{(k-1)!}}}\ds\big(a\big(\textstyle{1\over{k}}b^{k-1}+
\textstyle{1\over{k+1}}{\Sigma_{1, k-2}}
+\textstyle{1\over{k+2}}{\Sigma_{2, k-3}}+\ldots
+\textstyle{1\over{2k-1}}a^{2(k-1)}\big)\lin^0_1+$$
$$+\ts{{1\over{(k-1)!}}}\!\!\ds\int\limits^1_0\!\! dt
d\big(a\big(\textstyle{1\over{k}}{\Sigma_{\dot{1},0,
k-2}}+\textstyle{1\over{k+1}}{\Sigma_{\dot{1}, 1, k-3}}
+\textstyle{1\over{k+2}}{\Sigma_{\dot{1}, 2, k-4}}+\ldots
+\textstyle{1\over{2k-1}}{\Sigma_{\dot{1}, k-2, 0}}\big)\big)$$

}
 \vskip 5pt\noindent
The proof is based on the following lemma:
\vskip 5pt\noindent
{\bf Lemma 1.} {\sl The following identities hold in $V$:
$$\textstyle{{1\over{k}}d\big( a\Sigma[{\dot{a},(a^2)^0,
(b)^{k-2}}]\big)\treq{{k-1}\over{k}}\dot{a}b^{k-1}-{1\over{k}}a\overbrace{{b^{k-1}
}}^.}$$
$$
\textstyle{{1\over{k+1}}d\big(a\Sigma[{\dot{a},(a^2)^1,
(b)^{k-3}}]\big)\treq{{k}\over{k+1}}\dot{a}\Sigma[{(a^2)^1,
(b)^{k-2}}]-{{1}\over{k+1}}\dot{\Sigma}[{(a^2)^1,
(b)^{k-2}}]}$$
$$\textstyle{{1\over{k+2}}d\big(a\Sigma[{\dot{a},(a^2)^2,
(b)^{k-4}}]\big)\treq{{k+1}\over{k+2}}\dot{a}\Sigma[{(a^2)^2,
(b)^{k-3}}]-{{1}\over{k+2}}\dot{\Sigma}[{(a^2)^2,
(b)^{k-3}}]}$$
$$\hbox to 8cm{\dotfill}$$
$$\textstyle{{1\over{2k-1}}d\big(a\Sigma[{\dot{a},(a^2)^{k-2},
(b)^{0}}]\big)\treq{{2k-2}\over{2k-1}}\dot{a}(a^2)^{k-1}-
{{1}\over{2k-1}}\overbrace{(a^2)^{k-1}}^.
}$$
}
\vskip 5pt\noindent
{\bf Example 1.}
\vskip 5pt\noindent
\leftline{1. $k=2$}
 \vskip 5pt\noindent
$\ch_2^1(a(t))\treq {1  \over 2}
(ab+ {2  \over 3}a^3)\lin^1_0 +\ds{\int_0^1}dt \, d(\ts{1\over 2} a\dot{a})$

\vskip 5pt\noindent
 \leftline {2. $k=3$}
 \vskip 5pt\noindent
$\ch_3^1(a(t))\treq{1  \over 3!}(ab^2+ {3  \over 2}a^3b+{3  \over 5} a^5)
\lin^1_0 +\ds{1  \over 2!}\ds{\int_0^1dt }\, \ts d\big(a{1  \over 3}
(\dot{a}b+b\dot{a})+
{1  \over 4}(\dot{a}a^2+a^2\dot{a})\big)\treq$

\rightline{$\treq{1  \over 3!}(ab^2+ {3  \over 2}a^3b+{3  \over 5} a^5)
\lin^1_0 +\ds{1  \over 2}\ds{\int_0^1}dt \, \ts d\big({1  \over 3}
(a\dot{a}b-\dot{a}ab)+
{1  \over 2}a^3\dot{a}\big)$}
 \vskip 5pt\noindent
\leftline {3. $k=4$}
 \vskip 5pt\noindent
$\ch_4^1(a(t))\treq
{1  \over 4!}(ab^3+{8  \over 5} a^3bb+ {4  \over 5}
aba^2b+2a^5b+{4  \over 7} a^7)\lin^1_0 +$

$$+{1  \over 2!}\ds\int_0^1\!\!\! dt \,\ts d\big({1  \over 4}
(a\dot{a}-\dot{a}a)b^2+
{1  \over 4}ab\dot{a}b+{2  \over 5}a^3(\dot{a}b+b\dot{a})
+{1  \over 5} a\dot{a}a^2b+$$
$$\ts{1  \over 5} aba^2\dot{a}+{1  \over 2} a^5\dot{a}
\big).$$

\proof
\vskip 10pt\noindent
{\bf 6. Poisson bracket and  Chern characters.}
\vskip 5pt\noindent
{\bf Proposition 1} $  \{ \ch_k, \ch_l  \}  \treq 0$.
\vskip 5pt\noindent
{\bf Proof.} Let us notice that $  \partial_b (a^2+b)^r=r/\!\!\!b(a^2+b)^{r-1}$
That follows immediately from the definition of $  \partial_b$.
Also, $$  \partial_a
(a^2+b)^r=r(/\!\!\!aa(a^2+b)^{r-1}-/\!\!\!a(a^2+b)^{r-1}a)=
r(a(a^2+b)^{r-1}-(a^2+b)^{r-1}a).$$
\noindent
{\rm So} \ \  $\{ \ch_k, \ch_l  \} \treq  \{ (a^2+b)^k,(a^2+b)^l  \} \treq
  \partial_a (a^2+b)^k
 \partial_b (a^2+b)^l+$ $$ \partial_a (a^2+b)^l  \partial_b (a^2+b)^k\treq
k(a(a^2+b)^{k-1}-(a^2+b)^{k-1}a) l (a^2+b)^{l-1}+$$
\rightline{$+
 l(a(a^2+b)^{l-1}-(a^2+b)^{l-1}a) k(a(a^2+b)^{k-1})\treq$}
\vskip 3pt
\leftline{$\treq kl(a(a^2+b)^{k+l-2}-
(a^2+b)^{k-1}a (a^2+b)^{l-1})+$}
 $$ +kl( a (a^2+b)^{l-1}-
(a^2+b)^{k+l-2}  \cdot a(a^2+b)^{k-1}  \treq 0-0=0.\bx$$
\vskip 5pt\noindent
We have defined secondary characteristic classes $\ch_k^1$ in $V$ which
have the property $d\ch_k^1  \treq \ch_k$ and we have written an explicit
formula for them in the section 4. New let us look what
happens with their Poisson bracket and their bracket with ordinary Chern
characters.
\vskip 5pt\noindent
{\bf Proposition 2.} $d  \{ \ch_k^1, \ch_l  \} \treq 0$.
\vskip 5pt\noindent
{\bf Proof.} \  $$- d  \{ \ch_k^1, \ch_l  \}\treq  \{ d\ch_k^1,  \ch_l  \} +
(-1) ^{  \vert \ch_k^1  \vert}  \{ \ch_k^1,d \ch_l  \} \treq$$ $$ \treq \{
\ch_k,\ch_l
  \} -
 \{
\ch_k^1, 0  \} \treq0-0=0.\bx$$

\vskip 10pt\noindent
{\bf 7. Solving the equation $dP=Q$.}
\vskip 10pt \noindent
Let $P(a,b)$ be an element of a free associative algebra generated by 2
letters $a$ of degree 1 and $b$ of degree 2 the differential
$d$ that $da=b$, $db=0$.
 \vskip 5pt\noindent
Suppose $dP(a,b)=0$.
Let us take $$Q(a,b)= \int\limits_0^1 P(ta, dta+tb).$$
Then $dQ(a,b)=P(a,b)$. In the integral above we consider only
monomials with  $dt$ and we consider $dt$ as an element of degree
1 such that $ dt\cdot a=-a\cdot dt $,  $d\cdot tb=b
\cdot dt$  and  $dt  \cdot t= t  \cdot dt$.

 \vskip 5pt\noindent
{\bf Example 1.} Let $P(a,b)=ab-ba$. Then $ d(ab-ba)=b^2-b^2=0$.
$$Q(a,b)=  \int_0^1 ta (dta+tb)-(dta+tb) ta=  \int_0^1 (-tdt a^2 -
tdt a^2)=-a^2$$
$ d(-a^2)=-ba+ab.$

 \vskip 5pt\noindent
\exampleA

 \vskip 5pt\noindent

Now we can formulate a combinatorial algorithm for solving the
equation
$dQ=P$, where $dP=0$. Let $P=P_1+P_2+\ldots + P_l$, where $P_l$ are
monomials. For example, let $P_i=x_1  \ldots\ldots x_n,$
where $x_i$ one of the letters $a$ or $b$.
For each such monomials we construct a polynomial $Q_i$
\vskip 5pt\noindent

$\hskip 175pt a$

$\hskip 175pt \downarrow$

$Q_i=\displaystyle{{1  \over {  \| P_i  \|}}  \sum_{ over \ all \
x_k=b } (-1)^s  \underbrace{x_1  \ldots x_{k-1}}_{degree=s}} \
/\!\!\!{b}$ $x_{k+1}\ldots x_n$,

 \vskip 5pt\noindent
where $\|P_i\|$ is the number of letters in the word $P_i$.
\vskip 10pt\noindent
{\bf 8. Homotopy operators.}
\vskip 10pt\noindent
Let us construct now homotopy operators in the general case.
\vskip 5pt\noindent
{\bf Theorem 1.} Let $A$ be a free associative algebra generated by
elements $a_1, \ldots , a_n$ of degree 1 and elements $b_1, \ldots ,b_n$
of degree 2 with differential $d$ s.t. $da_i=b_i$  for all
$i=1, \ldots , n$.

Then for each   $P(a_1, \ldots ,a_n; b_1, \ldots ,b_n)$   the  homotopy
operator $h$ defined as $$hP(a_1, \ldots ,a_n; b_1, \ldots , b_n)=
 \int_0^1\!\!\! P(ta_1, \ldots ,ta_n; (dta_1+tb_1), \ldots ,(dta_n+tb_n))$$
satisfies the identity $P=d(hP)+h(dP)$ in particular if $dP=0$ then we can
solve equation $dQ=P$ taking $Q=hP$.
\proof
 \vskip 5pt\noindent
{\bf Theorem 1$'$.}   The following combinatorial formula holds for
$hP$. Let us write $P$ as a sum of monomials: $P=P_1+P_2+ \ldots +P_p$.
For example, $P_i=x_1  \ldots x_p$, where $x_i$ is one of $a$'s or
$b$'s
Then
\vskip 5pt\noindent

$\hskip 151pt a_{*}$

$\hskip 151pt \downarrow$

$hP_i=\displaystyle{{1  \over {  \| P_i  \|}}  \sum_{
x_k=b_{*} } (-1)^s  \underbrace{x_1  \ldots x_{k-1}}_{degree=s}} \,
/\!\!\!{b_*}$ $x_{k+1}\ldots x_p,$
\vskip 5pt\noindent
where
 $  \|P_i\|$ is
the total number of letters  in $P_i$. So we take a word $P_i$ of
 and replace each letter $b$ with some index *
by the letter $a$ with the same index. We give the resulting word the sign
$(-1)^s$, where $s$ is the sum of degrees of all letters in front of $b_{k}$.
So we get a sum of $\|P\|$
  words of degree $  \vert P_i  \vert -1$.
 \vskip 5pt\noindent
{\bf Example 1.} Let $P_i=baabab$. Then
\vskip 5pt\noindent
$Q_i=hP_i= {1  \over 6}
((-1)^0 aaabab +$ $ (-1)^4 baaaab +$ $ (-1)^7 baabaa +$
${ 1  \over 6}(aaabab + baaaab - baabaa)$.
\exampleB
 \vskip 5pt\noindent
{\bf Proposition 1.}
\vskip 5pt\noindent
1. The homotopy operator $h$ is a (co)differential in
algebra $A$:  $h  \circ h=0$.
\vskip 5pt\noindent
2. The operator $h$ is well-defined on the cyclic space $V$ and is
a
 (co) differential there: $h  \circ h  \lin_V  \treq 0$ and if $P  \treq 0$
then $hP  \treq 0$.
\vskip 5pt\noindent
3. Operators $h$ and $d$ in $A$ are connected by the formula
$h  \circ d + d  \circ h=id$;  so for every $Q  \in A$,  $$h(dQ) + d(hQ)=Q,$$
\vskip 5pt\noindent
4. The perators $h$ and $d$ in $V$ are connected by the formula
 $$h  \circ d + d  \circ h  \treq id$$ so for every $P  \in V$,
$$h(dP) + d(hP)  \treq P.$$
\proof
 \vskip 5pt\noindent
{\bf Example 2.} Let $P=baba  \treq 0$ because $baba=(-1)^{3  \cdot 3}baba$.
Then $hP={1  \over 4} (aaba + (-1)^3baaa)  \treq {1  \over 4} (baaa-baaa)=0$.
Let $M=baabab$. Then $$hM=\textstyle{1  \over 6} (aaabab + baaaab - baabaa)$$

$hhM={1  \over 6}({1  \over 6}(-aaaaab + aaabaa) + {1  \over 6}(aaaaab-baaaaa)
-{1  \over 6}(aaabaa+baaaaa))=0.\bx$
\vskip 5pt\noindent
Let us define now the
 operators ${h}_k$  for a monomial $P=x_1\ldots x_p$
 by the equality
\vskip 5pt\noindent
$\hskip 200pt a_k$
\vskip 0pt\noindent
$\hskip 200pt \downarrow$

$
 h_kP_i\defeq\displaystyle{{1  \over {  \| P_i  \|}}\!\!\!\!\!\!\!  \sum_{ over
\ all \
x_j=b_k} (-1)^s  \underbrace{x_1  \ldots x_{j-1}}_{degree=s}}
\, /\!\!\!{b_k}$ $x_{j+1}\ldots x_p, \hskip 10pt
h=\displaystyle{\sum\limits_{k=1}^N} h_k$
 \vskip 5pt\noindent
Define operators $\tilde{h}$, $\tilde{h}_k$
 for a monomial $P=x_1\ldots x_p$
 by the equality
\vskip 5pt\noindent
$\hskip 148pt a_{*}$
\vskip 0pt\noindent
$\hskip 148pt \downarrow$

{$\tilde{h}P=\!\!\!\!\!\! \displaystyle{\sum_{all \
 x_j=b_{*} }} (-1)^s  \underbrace{x_1  \ldots x_{j-1} }_{degree=s}
\, /\!\!\!{b_{*}}$ $x_{j+1}\ldots x_p,
 $}
\vskip 0pt\noindent
$\hskip 154pt a_{k}$
\vskip 0pt\noindent
$\hskip 154pt \downarrow$

$\tilde{h}_kP_i=\!\!\!\!\!\!\displaystyle{  \sum_{all \ x_j=
b_k} (-1)^s  \underbrace{x_1  \ldots x_{j-1}}_{degree=s}}
\, /\!\!\!{b_k}$ $x_{j+1}\ldots x_p,$

\vskip 5pt\noindent
Then $$(d\tilde{h}+\tilde{h}d)P=\|P\|P.$$
{\bf Proposition 2.}
 Operators $\tilde{h}$, $\tilde{h}_i$ and Poisson bracket are connected by
equalities
$$  -\tilde{h}  \{ P,Q  \} =  \{ \tilde{h}P, Q  \} + (-1)^{ \vert P  \vert}  \{
P, \tilde{h}Q  \}-2\sum_{i
=1}^N\partial_{b_i}P\cdot \partial_{b_i}Q,\ $$
$$  -\tilde{h}_i  \{ P,Q  \} =  \{ \tilde{h}_iP, Q  \} +
 (-1)^{ \vert P  \vert}  \{\
 P, \tilde{h}_iQ  \}-2\partial_{b_i}P\cdot \partial_{b_i}Q,\ $$
The proof is based on the following lemma.
\vskip 5pt\noindent
{\bf Lemma 1.}{\sl \  The following equalities are true in $A$}
 $$\partial_{b_i}\circ \tilde{h_i}=\tilde{h_i}\circ \partial_{b_i},$$
$$\partial_{a_i}
\circ \tilde{h_i}=-\tilde{h_i}
\circ \partial_{a_i}+\partial_{b_i}.$$
\proof
 \vskip 5pt\noindent
{\bf Proposition 3.} In the cyclic space $V$ operators $h$ and
$\tilde{h}$ can be written as
$$hP  \treq {1  \over {  \| P\|}}  \sum\limits_{i=1}^N a_i ({ \partial\over
\partial{b_i}}P), \  \ \  \ \ \tilde{h}P  \treq  \sum\limits_{i=1}^N a_i ({
\partial\over
\partial{b_i}}P).$$
 \vskip 5pt\noindent
{\bf Theorem 2. \ } {\sl\  Operators $d_i$ and $h_i$, $i=1,\ldots, N$ satisfy
following identities:
\vskip 10pt\noindent
${(1)}\hskip 15pt dh+hd=id, \hbox{\sl \  where \ }
d=\displaystyle{\sum\limits_i d_i,
\hbox{\sl \ and \ } h=\sum\limits_i h_i}$
\vskip 4pt\noindent
${(2)}\hskip 15pt d_id_j+d_jd_i=0,$
\vskip 10pt\noindent
${(3)}\hskip 15pt h_ih_j+h_jh_i=0,$
\vskip 10pt\noindent
${(4)}\hskip 15pt h_id_j+h_jd_i=0 \hbox{ \sl \ for \ } i\neq j.$
\vskip 10pt\noindent
${(5)}\hskip 15pt \hbox{\sl For an arbitrary monomial \ } P\in A$

\vskip5pt
$\hskip 10pt \hskip 15pt  (d_ih_i+h_id_i)P={{\|P\|_{i}}\over \|P\| }P,$
\vskip 5pt\noindent
 where  ${\|P\|_{i}}$ is the number of letters
$a_i$ and $b_i$ in a monomial $P$,
 and $\|P\|$ is the total number of letters in
$P$. $\|P\|$ should not be confused with $|P|$ which is the total degree
of $P$ i.e. sum of degrees of all letters in $P$.}
\proof
 \vskip 10pt\noindent
{\bf 9. Action of the Gauge group and its algebraization.}
\vskip 10 pt\noindent
We repeat  here
 the  standard basic facts about connections and
curvature which we are going to algebraize in the next section.
Consider a vector bundle $E$ over a base manifold $M$.
Let
$$ [e]=  \pmatrix{e_1(x)  \cr  \vdots  \cr e_n(x)} $$
be a basis of local sections
of $E$. Let $$  \pmatrix {e_1  '(x)  \cr  \vdots  \cr e_n  '(x)}=  \pmatrix
{ \tilde{\sigma}_1^1(x) &  \tilde{\sigma}_1^2(x) & \ldots   \cr
\tilde{\sigma}_2^1(x)   \cr  \vdots & \ldots
&  \vdots}  \pmatrix {e_1(x)  \cr  \vdots  \cr e_n(x)}$$ be another basis.
We shall write $[\cdot ]$ for a column and $(\cdot)$ for a row.
Let $(f)=(f_1(x) \ldots  f(x))$ be coordinates of the section of the
bundle written in the basis $[e]$. Then the section $f$ can be written as
$f=(f)[e]$. We define
 a connection  $ \nabla$ as an operator on sections satisfying

$$  \nabla (  \mu f)=d  \mu  \cdot f +  \mu  \nabla f, $$ for an arbitrary
function
 $ \mu$
and  arbitrary section $f$  of $E$.
A connection acts on the basic sections as
$$ \nabla  \pmatrix{e_1  \cr \vdots  \cr e_n}=  \pmatrix { \tilde{\omega}_1^1
&  \tilde{\omega}_2^1
& \ldots   \cr  \tilde{\omega}_1^2 & &  \vdots  \cr  \vdots & \ldots  &
\vdots}
 \pmatrix {e_1  \cr  \vdots  \cr  e_n}=  \tilde{\omega}[e].$$
And the matrix $ \tilde{\omega}$ of 1-forms which defines  the connection
 transforms as follows:
$$  \tilde{\omega}  \to d  \tilde{\sigma}  \cdot  \tilde{\sigma}^{-1}+
\tilde{\sigma}  \tilde{\omega}  \tilde{\sigma}^{-1}.$$
$$R=^t(d  \tilde{\omega} -  \tilde{\omega}  \wedge
 \tilde{\omega})=d{\omega} + {\omega}  \wedge {\omega}$$
is called the curvature  tensor. Here $\omega$ is  the matrix
of 1-forms which is the transposed of $\tilde{\omega}$.

A
connection $\nabla$ can be written as $ d+\omega$:
$$ \nabla [f] = d[f]+\omega[f] \hbox { \ \ \  and \ \ \ \ \ }
  \nabla (f)=d(f)+(f)\tilde{\omega}.$$
Consider now the gauge transformation $  \tilde{\sigma}$. It transforms
$\omega$:
$$\omega  \to  \sigma^{-1}d  \sigma +   \sigma^{-1} \omega  \sigma,$$
where $\sigma$ is the transposed of $\tilde{\sigma}$.

\vskip 10pt\noindent
{\bf 10. Gauge  algebra and secondary classes
.}
\vskip 10 pt\noindent
Let us correspond
$$\matrix{\hbox{to}&\omega_i& \  \  \   \ & a_i,\cr
\hbox{to}& d\omega_i& \  \ \ \   \ & b_i,\cr
 \hbox{to}&\sigma^{-1}& \  \ \ \   \ & y,\cr
\hbox{to}&\sigma& \  \ \   \ & x,\cr
\hbox{to}&d\sigma^{-1}& \  \ \ \   \ & q,\cr
\hbox{to}&d\sigma& \  \ \ \   \ & p.\cr}$$
Consider an associative algebra $G$ generated by letters $a_i,
b_i,$ $x,\  y,\  p,\  q$, \  $i=1\ldots N$
with a differential $d$ such that

\centerline{$da_i=b_i,$}

\centerline{$db_i=0,$}

\centerline{$dx=p,$}

\centerline{$dp=0,$}

\centerline{$dy=q,$}

\centerline{$dq=0.$}
\noindent
subject to the relation
{$xy=yx=1.$} We shall call $G$ {\it gauge algebra}.
\noindent
 From the equation $xy=yx=1$ we get
$$d(xy)=d(yx)=0,$$ $$py+xq=0, \ \ qx+yp=0.$$
 From the equation  $py+xq=0$ we obtain $$ {-p=xqx, \ \   -q=ypy}.$$
We can obtain a whole series of  identities by  further differentiating
  $xy+yx=1$. We shall make much use of the expression $py$ and we introduce
a special letter $c$ for it.
$$dc=c^2,$$
because $dc=d(py)=-pq=-p(-ypy)=(py)^2$.

Elements of the  algebra $G$ have the following degrees:
 $$  x, \ y  \  \hbox{\rm \ are of  \ degree  \ 0,}$$
  $$ p, \ q, \ a_i \  \hbox{\rm \  are of  \ degree  \ 1,}$$
 $$ b_i \  \hbox{\rm \ are of \ degree 2,}$$
$$ c \  \hbox{\rm \ is of \ degree 1.}$$
 The gauge transformation

$a  \to yp + yax=y(a+c)x$  transforms

\vskip 5pt\noindent

$b=da \to dyp+ydp+dyax+ydax-yadx=y(b+a^2-(a+c)^2)x$.
\noindent
We shall  consider a cyclic
space $V_G$ for $G$: $V_G=G / [G,G]$. We shall write $\treq$ for equality
in $V_G$.
\vskip 5pt\noindent
{\bf Example 1.}

$b \to dyp+ydp+dyax+ydax-yadx=$

\rightline{$=qp+qax+ybx-yap=ybx-yap+qax+qp$}

\vskip 5pt\noindent

$a^2  \to (yp+yax)(yp+yax)=ypyp+yap+ypyax+ya^2x=$

\vskip 0pt \noindent $
ya^2x+yap-qax-qp.$

 \vskip 5pt\noindent

 $b+a^2  \to y(b+a^2)x. $
\vskip 5pt\noindent
\leftline{ So $a^2+b$ behaves as  a tensor under the gauge transformation.}

\vskip 5pt\noindent
{\bf Example 2.}
 Let us look at what happens to the element
$\ch_2^1=\textstyle{1\over 2}(ab+\textstyle{2\over 3}a^3)$ under the gauge
transformation.

$$\textstyle{1\over 2}(ab+\textstyle{2\over 3}a^3)
\to \textstyle{1\over 2}(ab+\textstyle{2\over 3}a^3)+\textstyle{1\over
2}(cb-c^2a-
\textstyle{1\over 3}c^3).$$
We can check explicitly that $$d\textstyle{1\over 2}(cb-c^2a-\textstyle{1\over
3}c^3)\treq0.$$
We know $da=b$, $db=0$, $dc=c^2$.
$$d\textstyle{1\over 2}(cb-c^2a-\textstyle{1\over 3}c^3)=\textstyle{1\over
2}(c^2b-c^2ca+cc^2a-c^2b-\textstyle{1\over 3}c^4)\treq0,$$ because $c^4\treq
cc^3\treq-c^3 c\treq 0.$
\vskip 5pt\noindent
{\bf Theorem 1.} {\sl Under the gauge transformation in the cyclic space
$V(G)$ the Chern-Simons class $\ch_k^1(a)$  transforms to
$$\ch_k^1(a)\to\tilde{\ch_k^1}\treq \ts{1\over (k-1)!}\ds(a+c)\!\!\!
\!\!\!\!\!\sum_{\alpha+\beta+\gamma=k-1}\!\!\!\!\!\!\!(-1)^\gamma
{{(k+\beta-1)!\gamma
!}\over (k+\beta+\gamma)!}\Sigma[(b)^\alpha, (a^2)^\beta, (u)^\gamma],$$
where $u=ca+ac+c^2$ and $\Sigma[(b)^\alpha, (a^2)^\beta, (u)^\gamma]$
is the sum of all possible words in $G$ with $\alpha$ letters
$b$, $\beta$ letters $a^2$ and $\gamma$ letters $u$.}
\vskip 5pt\noindent
{\bf Proof.} $$\tilde{\ch_k^1}={1\over k!}\int^1_0dt k(a+c)(tb+t^2a^2+(t^2-t)
(ca+ac+c^2))^{k-1}=$$
$$={1\over (k-1)!}\int^1_0dt (a+c)(tb+t^2a^2+(t^2-t)u)^{k-1}=$$ $$
{1\over (k-1)!}\sum_{\alpha+\beta+\gamma=k-1}\int^1_0
dt\, t^\alpha t^{2\beta}(t^2-t)^\gamma
\Sigma[(b)^\alpha, (a^2)^\beta, (u)^\gamma]=$$
$$={1\over (k-1)!}(a+c)\sum_{\alpha+\beta+\gamma=k-1}(-1)^\gamma
{{(k+\beta-1)!\gamma
!}\over (k+\beta+\gamma)!}\Sigma[(b)^\alpha, (a^2)^\beta, (u)^\gamma],$$
because $$\int^1_0\!\!\!\!\!dt\, t^{\alpha+2\beta+\gamma}(t-1)^\gamma=
(-1)^\gamma\!\!\!\!\int^1_0\!\!\!\!\!dt\, t^{k-1+\beta}(t-1)^\gamma
=(-1)^\gamma{{\Gamma(k+\beta)\Gamma(\gamma+1)}\over\Gamma(k+\gamma+\beta+1)}.\bx$$
\vskip 5pt\noindent
{\bf Example 3.} Under the gauge transformation secondary classes
$\ch_2^1(a)$ and $\ch_3^1(a)$ are transformed in the following way.
To translate these formulas to the most common notation
we should take $a=A$ matrix of 1-forms defining the connection
take $b=dA$ matrix of two forms. For a
 gauge transformation $g:A\to g^{-1}dg+g^{-1}Ag$ we take
$c=dg\cdot g^{-1}$. We should also  put $Tr$ in front of the whole expression.
\vskip 5pt\noindent
$$\textstyle{\ch_2^1(a)\to\tilde{\ch_2^1}\treq{1\over {1!}}(a+c)({1\over 2}b +
{1\over 3}a^2-{1\over 6}(ca+ac+c^2))\treq}$$
$$\treq \textstyle{1\over 2}(ab+\textstyle{2\over 3}a^3)+
\textstyle{1\over 2}(cb-c^2a-\textstyle{1\over 3}c^3).$$
So $\textstyle{1\over 2}Tr(AdA+\textstyle{2\over 3}A^3)\to\textstyle{1\over
2}Tr(AdA+\textstyle{2\over 3}A^3)+
\textstyle{1\over 2}((dg\cdot g^{-1}) dA-(dg\cdot g^{-1})^2A-$
$$\textstyle{1\over 3}(dg\cdot g^{-1})^3)$$
$$\textstyle{
\ch_3^1(a)\to\tilde{\ch_3^1}\treq{1\over {2!}}(a+c)({1\over 3}b^2 +
{1\over 4}(ba^2+a^2b)+{1\over 5}a^4-{1\over 12}(bu+ub)+}$$
$$\textstyle{+{1\over 30}u^2-{1\over 20}(a^2u+ua^2))+)}\treq$$
$$\textstyle{ \treq{1  \over 3!}(ab^2+ {3  \over 2}a^3b+{3  \over 5} a^5)+
{1  \over 3!}(
{1\over 10}c^5+{1\over 2}ac^4-{1\over 2}c^3b+cb^2-{1\over 2}c^2a^3+
{1\over 2}aca^2c}$$ $$-\textstyle{{1\over 2}[cca+cac+acc]b+{1\over 2}[caa
-aca+aac])}$$
The  expression ${1\over 3!}({1\over 10}c^5+{1\over 2}ac^4 - ...)$
 is $d$ closed in the cyclic space $V_G$.
Let us check it.
$$d\big(\textstyle{
{1\over 10}c^5+{1\over 2}ac^4-{1\over 2}c^3b+cb^2-{1\over 2}c^2a^3+{1\over
2}ac\
a^2c}-$$ $$-\textstyle{
{1\over 2}[cca+cac+acc]b+{1\over 2}[caa-aca+aac]b}\big)\treq$$
$$\textstyle{\treq{1\over 10}c^6+{1\over 2}bc^4-{1\over 2}c^4b+c^2b^2-
{1\over 2}c^2(baa-aba+aab)+}$$ $$\textstyle{+{1\over 2}
(bcac^2-ac^2ac^2+acbc^2)-}$$ $$-\textstyle{
{1\over 2}[ccb-cbc+c^2ac+cac^2+bcc]b+{1\over 2}[c^2a^2+ac^2a
+a^2c^2]b\treq 0.}$$
\vskip 5pt\noindent
{\bf Example 4.} {\footnote{${}^*)$}{We want to thank Lisa Jeffrey who
brought to our attention
this example.}}
Let $M$ be a 3-manifold and $g:M\to G=SU(2)$ is a smooth mapping.
Let $\lambda\in\Omega^3(G)$ be a fundamental form
which generates $H^3(G)$ . Then
$$\deg( g)=\int_M g^*\lambda.$$
Consider a principal $SU(2)$ bundle over $M$  with connection $A\in
\Omega^3(M)\otimes\hbox{\cal g}$. Define
$$CS(A)=\int_MTr(AdA+\textstyle{2\over 3}A^3).$$
The
 gauge transformation acts on $A$ \ $g:A\to A^g=g^{-1}dg+g^{-1}Ag$.
$$CS(A^g)=\int_MTr((A^g)d(A^g)+\textstyle{2\over 3}{(A^g)}^3)+8\pi^2\deg(g).$$
But from calculations in our algebra we know that
$$\textstyle{\ch_2^1(a)\to\tilde{\ch_2^1}\treq{1\over {1!}}(a+c)({1\over 2}b +
{1\over 3}a^2-{1\over 6}(ca+ac+c^2))\treq} \ \ \ \hbox{i.e.}$$
$$Tr(AdA+\textstyle{2\over 3}A^3)\!\to\!
Tr(AdA+\textstyle{2\over 3}A^3)+
\textstyle{}((dg\cdot g^{-1}) dA-(dg\cdot g^{-1})^2\!A-
\textstyle{1\over 3}(dg\cdot g^{-1})^3),$$ so
$$\deg(g)={1\over 8\pi^2}\int_M Tr((dg\cdot g^{-1}) dA-(dg\cdot g^{-1})^2A-
\textstyle{{1\over 3}(dg\cdot g^{-1})^3)}=\displaystyle{\int_Mg^*\lambda.}$$





\vskip 10pt\noindent
{\bf 11. Appendix. Explicit formulas for secondary classes $\ch_k^2$.}
\vskip 10pt\noindent
Let $\Delta$ be a 2-simplex
$t_0+t_1+t_2=1,$ $ t_0, t_1,
t_2>0$.
Consider 3 connections $a_0$, $a_1$, $a_2$ on the bundle $E\to M$.
Consider the connection $a(t)$, $t\in \Delta$ such that
$a(t)=t_0a_0+t_1a_1+t_2a_2$. Then
     $$\ch_k^2={1\over k!} \tr \int_\Delta
       [d(t_0a_0+t_1a_1+t_2a_2)+(t_0a_0+t_1a_1+t_2a_2)^2]^k=$$ $$
       ={1\over k!}   \int_\Delta
[(dt_0a_0+dt_1a_1+dt_2a_2)+(t_0b_0+t_1b_1+t_2b_2)+
       (t_0a_0+t_1a_1+t_2a_2)^2]^k=$$
 $${1\over k!}  \tr \int_\Delta (P+Q)^k,$$
 where      {$$P=dt_0a_0+dt_1a_1+dt_2a_2,$$  $$Q=(t_0b_0+t_1b_1+t_2b_2)+
       (t_0a_0+t_1a_1+t_2a_2)^2$$ and both $P$ and $Q$ are of degree 2.}
In the calculations below we shall consider $a_i$ and $b_i$ as
the elements of the cyclic space $V $
and we shall omit $\tr$ sign in equalities.
 We are interested only in summands with 2 letters $P$ and $(k-2)$
letters $Q$ because only they will give a 2-form in $dt$ which can be
can be integrated over the 2-simplex $\Delta$. Let us take
           $t_0=1-t_1-t_2$, \ then $dt_0=-(dt_1+dt_2)$.
When integrating over $\Delta$ we in fact
 integrate over the triangle $0\leq t_1\leq 1$, $0\leq t_2\leq 1- t_1$.
 $$\int_\Delta (\ldots)\defeq\int_\Delta dt_1dt_2(\ldots)
=\int_{0}^{1}dt_1\int_{0}^{1-t_1}dt_2(\ldots)
.$$ We have
$$P^2=(dt_0a_0+dt_1a_1+dt_2a_2)^2=$$ $$=dt_1dt_2[a_1, a_0]+dt_1dt_2[a_0,
a_2]+dt_1dt_2[a_2, a_1]=$$ $$=dt_1dt_2([a_
      1, a_0]+[a_0, a_2]+[a_2, a_1]).$$
For future calculations  we shall use the following integral:
$$ \int_\Delta t_0^ \alpha t_1^ \beta t_2^ \gamma\, dt_1dt_2= { \alpha !
     \,   \beta ! \, \gamma ! \over (\alpha + \beta + \gamma
       +2)!}$$
{\bf Calculation of $\ch_k^2(a_0, a_1, a_2)$.}
\vskip 5pt\noindent
Case $k=2,$ \ $$\ch_2^2(a_0, a_1, a_2)={1\over 2!}  \int_\Delta PP
= (a_1a_0+a_0a_2+a_2a_1)={1\over 2}(-A_1A_2+A_2A_1),$$
where $A_1=a_1-a_0, A_2=a_2-a_0.$

       Case $k=3,$  $$\ch_3^2(a_0, a_1, a_2)=
 {1\over {3!}}   \int_\Delta (PPQ+PQP+QPP)= {1\over {3!}}  \ 3  \int_\Delta
PPQ=$$
$$={1\over {3!}}3([a_1,a_0]+[a_0,a_2]+[a_2,a_1])\big({1 \over 6} (b_0+b_1+b_2)+
$$
$$+{1 \over 12}((a_0^2 + a_1^2 +
       a_2^2) + {1 \over 24}( (a_0a_1+a_1a_0) + (a_1a_2+a_2a_1) + (a_0
       a_2+a_2a_0) )\big).$$
Really
 $$\int_\Delta(t_0b_0+t_1b_1+t_2b_2)(t_0a_0+t_1a_1+t_2a_2)^2=
{1\over 6}(b_0+b_1+b_2)+$$ $$
\int_\Delta (t_0^2a_0^2+t_1^2a_1^2+t_2^2a_2^2)+
      \int_\Delta ( t_0t_1\Sigma [a_0, a_1] + t_1t_2\Sigma [a_1, a_2] +
 t_0t_2\Sigma [a_0, a_2])=
      $$ $$ ={1\over 6}(b_0+b_1+b_2)+{1 \over 12}(a_0^2 + a_1^2 + a_2^2) + {1
\over 24} (\Sigma [a_0, a_1]\
 +       \Sigma [a_1, a_2] + \Sigma [a_0, a_2]),$$ because
$$\int_\Delta t_i={1! \over 3!}={1 \over 6}, \ \ \  \int_\Delta t_i^2
={2! \over 4!}={1 \over 12}, \ \ \
 \int_\Delta t_it_j={1! \over 4!}={1 \over 24},$$
where $i, j=1,2,3$ and $i\neq j$.

      When $k=3$ and $a_2=0$
       $$\ch_3^2(a_0, a_1, 0)={1\over {3!}}
\int( \ldots  )^3 = $$ $$={1\over {3!}}3[a_1, a_0](
 {1 \over 12} (a_0^2 + a_1^2) + {1 \over\
 24}
       (a_0a_1 + a_1a_0) + {1 \over 6} (b_0 + b_1))
=$$
$$={1\over {3!}}{ 3 \over 6} (a_1a_0 -
       a_0a_1)( (b_0+b_1)+ {1 \over 2} (a_0^2+a_1^2+ {1 \over 2} (a_0a_1 +
       a_1a_0)))$$
We shall publish the results of other calculations elsewhere.
 \vskip 10pt\noindent
{\bf 12. Appendix. Formulas for a  secondary class
$\ch_k^1(a_0, a_1)$}
\vskip 10pt\noindent
Now let us write a  formula for a secondary class
$$\ch_k^1(a_0, a_1)\defeq{1\over k!}\int_0^1((dt_0a_0+dt_1a_1)+(t_0b_0+t_1b_1)
+(t_0a_0+t_1a_1)^2).$$
\vskip 5pt\noindent
{\bf Theorem.}{\sl
$$ \ch_k^1(a_0, a_1)= {1 \over (k-1)!} A \sum_{m+s+r=k-1}
 {1 \over 2m+s+1}\Sigma_{(m,s,r)}$$

where       $A=a_1-a_0$, \ \ $R=b_0+a_0^2$,\  \ $S=(b_1-b_0)
+(a_0A+Aa_0)$,\ \ $M=A^2$    \ and \  $ \Sigma_{(m,s,r)}$ is the sum of all
 possible words with $m$ letters  $M$,
       $s$ letters $S$ and $r$ letters $R$. }
\vskip 5pt\noindent
{\bf Sketch of the proof.}

$$ \ch_k^1(a_0, a_1)={1\over k!}\int_0^1((dt_0a_0+dt_1a_1)+(t_0b_0+t_1b_1)
+(t_0a_0+t_1a_1)^2)=$$
$$={1\over k!}\!\int\limits_0^1 (dt_1\,\underbrace{(a_1\!-a_0)}_{\hbox{\it A}}
+\underbrace{(b_0+a_0^2)}_{\hbox{\it R}}+$$
$$+t_1\underbrace{[(b_1-b_0)
+(a_0(a_1-a_0)+(a_1-a_0)a_0)]}_{\hbox{\it S}}\!+t_1^2\!\underbrace
{(a_1-a_0)^2}_{\hbox{\it M}})^k$$
$$={1\over k!}\int_0^1 (dt\,A +R+tS +t^2M )^k={1\over k!}\int_0^1 kdt\,A (R+tS
+t^2M )^{k-1}=$$
$$ ={1 \over k!} kA \sum_{m+s+r=k-1}
 {1 \over 2m+s+1}\Sigma_{(m,s,r)}.\bx$$
\vskip 5pt\noindent

  \vskip 15pt
\centerline{REFERENCES}
\vskip 5pt\noindent

\item{[GGL]}
 A. Gabrielov, I.M. Gelfand, and M. Losik,
{\it Combinatorial
calculation of characteristic class, }
 Funktsional Anal. i Prilozhen.( Funct. Anal. Appl.) vol 9
 1975
 54--55
 no. 2, 12--28
 no. 3, 5--26
\vskip 1pt
\item {[GFK]} Gelfand I.M., Fuks, D.B., Kazhdan, D.A.,{\it The actions of
infinite
dimensional Lie algebras}, Funct. Anal, Appl, {\bf 6} (1972) 9-13.

\vskip 1pt

\item{[GTs]}
I.M. Gelfand and B.L.Tsygan, {\it On the localization
of topological invariants,} Comm. Math. Phys., 146 1992 73-90
\vskip 1pt

\item{[M]}
 R. MacPherson, {\it The combinatorial formula of
Garielov, Gelfand,
and Losik for the first Pontrjagin class, }
 S\'eminaire Bourbaki No. 497
 Lecture Notes in Math., vol. 667
 Springer,
 Heidelberg

 1977
\vskip 1pt

\item{[GM]}
I.M. Gelfand and R. MacPherson, {\it A combinatorial formula
for the  Pontrjagin classes, }
Bull AMS vol.26 no.2 1992 304-309

\vskip 1pt

\item{[M]}
 R. MacPherson, {\it Combinatorial differential manifolds,} in
 Topological methods in modern mathematics : a symposium in
                   honor of John Milnor's sixtieth birthday, Lisa
                   R. Goldberg and Anthony V. Phillips, eds.,
 Publish or Perish, Inc., Houston,  1993.

\vskip 1pt

\item{[Kon]} M.Kontsevich,{\it Formal (Non)-commutative symplectic geometry,}
{\it in}
The Gelfand Mathematical Seminars 1990-92, L.Corwin, I.Gelfand, J.Le\-pow\-sky
eds.,
 Birkh\"auser, Boston 1993, 173-189.
\vskip 1pt

\item{[CS]} S.S.Chern, J.Simons, {\it
 Characteristic forms and geometric invariants,} Ann. Math, vol. 99
1974 48-69
\vskip 1pt

\item{[B]}
R.Bott, {\it On the Chern-Weil homomorphism and
the continuous cohomology of Lie groups,} Advances in Math,
vol. 11 1973  289-303
\vskip 1pt

\item{[BSS]} R.Bott, H.Shulman, J.Stasheff, {\it On the de Rham theory
of certain classifying spaces,}  Advances in Math,
vol. 20 1976 43-56
\vskip 1pt

\item{[C]}
 J. Cheeger, {\it Spectral geometry of singular
Riemannian spaces,}
 J. Differential Geom., vol 18  1983
 575--657
\vskip 1pt

\item{[Q]} D.Quillen, {\it Superconnections and the Chern character,}
Topology, vol. 24  1985 no.1 89-95
\vskip 1pt

\item{[MQ]} V.Mathai, D.Quillen, {\it Superconnections, Thom classes, and
equivariant differential forms, } Topology vol.25 1986 85-110
\vskip 1pt

\item{[LQ]} J.L.Loday, D.Quillen,  {\it Cyclic homology
and the Lie algebra of matrices, } Topology vol.24 1985 89-95

\vskip 1pt

\item{[LQ]} J.L.Loday,   { Cyclic homology, } Springer, Berlin 1992.
\vskip 1pt
\item{[Du]} J. Dupont, {\it The Dilogarithm as a characteristic class of a flat
bundle,}
J.Pure and Appl. Algebra, vol.44 1987 137-164.
\vskip 1pt

\vskip 1pt
\item {[O]} Olver, P.J., {\it Applications of Lie groups to the partial
differential equations}, Springer, Berlin 1985.

\vskip 1pt
\item {[Ts]} Tsujishita, T., {\it On variation bicomplex associated to
differential equations}, Osaka J. Math, {\bf  19} (1982) 311-363.
\vskip 1pt
\item {[KLV]} Krasil'shchik, I.S., Lychagin, V.V., Vinogradov, A.M., {\it
Geometry
of jet spaces and nonlinear partial differential equations}, Gordon and Breach,
NY 1986.
\vskip 1pt
\item {[BR]} Beresin, F., Retakh, V.,{\it A Method of computing characteristic
classes
of vector bundles,} Reports on Mathematical Physics, {\bf 18} (1980) 363-378

\vskip 1pt
\item {[LZ]} Lian B.H., Zuckerman, G.J.{\it
New perspectives on the BRST-algebraic structure of string theory,}
hep-th / 9211072
\vskip 1pt
\item {[GFK]} Positselskii, L.,{\it Quadratic duality
and curvature,
 } Funct. Anal. Appl., {\bf 27} (1993) 197-204.

\end